\documentclass[doublecol]{epl2}
\usepackage{graphics}
\usepackage{graphicx}% Include figure files
\usepackage{dcolumn} % Align table columns on decimal point
\usepackage{bm}
\usepackage{color}
\usepackage{epsfig}

\usepackage{amsmath}
\usepackage{subfigure,dcolumn}
\usepackage{multirow,longtable,textgreek,upgreek}

\title{Doping-Induced Spectral Shifts in Two Dimensional Metal Oxides}
\author{E. R. Ylvisaker and W. E. Pickett}
\institute{Department of Physics, University of California Davis, Davis, California 95616}

\date{\today}


\abstract{Doping of strongly layered ionic oxides is an established paradigm for creating
novel electronic behavior. This is nowhere more apparent than in superconductivity,
where doping gives rise to high temperature superconductivity in cuprates (hole-doped)
and to surprisingly high T$_c$ in HfNCl (T$_c$=25.5K, electron-doped). 
First principles calculations of hole-doping of the layered delafossite
CuAlO$_2$ reveal unexpectedly
large doping-induced shifts in spectral density, strongly in opposition to the
rigid band picture that is widely used as an accepted guideline.  These spectral 
shifts, of similar origin as the charge transfer used to produce negative
electron affinity surfaces and adjust Schottky barrier heights, 
drastically alter the character of the Fermi level 
carriers, leading in this
material to an O-Cu-O molecule-based carrier (or polaron, at low doping)
rather than a nearly pure-Cu hole as in a
rigid band picture. 
First principles linear response electron-phonon coupling (EPC) 
calculations reveal, as a consequence, net {\it weak EPC} and no superconductivity
rather than the high T$_c$ obtained previously using rigid band expectations.  
These specifically two-dimensional dipole-layer driven spectral shifts provides new insights into 
materials design in layered materials foe functionalities besides superconductivity.}
\begin{document}
\maketitle 

%\section{Background}

The quest for new materials functionalities is especially vigorous in transition
metal oxides (TMOs), with quasi-two dimensional (q2D) classes causing great
activity. The cuprate superconductors, with high superconducting critical temperature (HTS)
T$_c$, provide the most prominent example, but doping-induced superconductivity arises in
numerous other unexpected systems: ${\cal M}$NCl, ${\cal M}$ = Ti, Zr, Hf (T$_c$=15-25K); 
MgB$_2$, a self-hole-doped superconductor at 40K; the triangular lattice oxides
Li$_x$NbO$_2$, Na$_x$CoO$_2$, and chalcogenides Cu$_x$TiSeO$_2$ and 
A$_x$${\cal T}$S$_2$ (A=alkali, ${\cal T}$=transition metal), all 
with\cite{other} T$_c \sim$ 5K. 
The cuprates, followed
by MgB$_2$ and then by the Fe pnictide superconductors (FeSCs) 
with T$_c$ up to 56 K,, have illustrated that excellent 
superconductors appear in surprising regions of the materials palette. Even the FeSCs can
be pictured as doped (or self-doped) semimetallic superconductors.
 
The CuO$_2$ square-lattice cuprates have inspired study -- the computational design -- of related
square-lattice transition metal oxides, such as the ``charge conjugate''
vanadate\cite{sr2vo4A,sr2vo4B,sr2vo4b2} Sr$_2$VO$_4$, the Ag$^{2+}$ 
material\cite{deepa}  Cs$_2$AgF$_4$
that is isostructural and isovalent with La$_2$CuO$_4$, and cuprate-spoofing 
artificially layered nickelates,\cite{lanio3A} 
so far without finding new superconductors.\cite{sr2vo4C,lanio3B}
These highly interesting materials, though unfruitful for their original intent,
suggests that a more detailed understanding of doping effects is necessary
to unravel the mechanism of pairing in q2D systems.
Nevertheless, materials
property design can and does proceed when there is some broad understanding of 
the mechanism underlying the property.\cite{spaldin,lehur,pardo,negU,blundell}  

The superconducting pairing mechanism is only well understood for electron-phonon coupling (EPC)
where MgB$_2$ with T$_c$=40 K is most successful so far. 
The detailed understanding of EPC through strong-coupling 
Eliashberg theory\cite{SSW,gunnarsson}
encourages rational, specific optimization of the EPC strength $\lambda$ and of T$_c$, 
and specific guidelines for one direction for increasing T$_c$ 
have been laid out.\cite{RTS}

Recently a new and different class of cuprate,
the delafossite structure CuAlO$_2$ $\equiv$ AlCuO$_2$, has been predicted 
by Nakanishi and Katayama-Yoshida\cite{NK-Y} (NK)
to be a T$_c \approx$ 50K
superconductor when sufficiently hole doped.  The calculated EPC strength and
character is reminiscent of that of
MgB$_2$, whose high T$_c$ derives from a specific mode (O-Cu-O stretch for CuAlO$_2$) and focusing in 
$q$-space\cite{JAn,OKA,IIM,Bohnen}
due to the circular shape of the quasi-two dimensional (2D) Fermi surface (FS).
CuAlO$_2$ is another layered cuprate, with Cu being (twofold) coordinated
by O ions in a layered crystal structure. The differences with square-lattice
cuprates are however considerable:
the Cu sublattice is not square but triangular; there are {\it only} apical oxygen neighbors;
the undoped compound
is a $d^{10}$ band insulator rather than a $d^9$ antiferromagnetic
Mott insulator; it is nonmagnetic even 
when lightly doped; and it is most heavily studied as a $p$-type transparent conductor.\cite{transparent}
It shares with the hexagonal ${\cal M}$NCl system 
that doped-in carriers enter at a $d$ band edge. 
NK provided computational evidence for
impressively large $\lambda$
and high temperature superconductivity
T$_c$ up to 50 K when this compound is hole-doped, {\it viz.} CuAl$_{1-x}$Mg$_x$O$_2$.  
It is known that the delafossite structure is retained at least to $x$=0.05 upo
coping with Mg.\cite{MgDoping}
If this prediction could be substantiated, a new and distinctive structural class would be opened
up for a more concerted search for high temperature superconductors (HTS).

When our initial linear response calculations indicated weak (rather than strong)
EPC, we performed a
more comprehensive study. 
In their work, NK did not carry out linear response calculations of electron-phonon
coupling for doped CuAlO$_2$.  Instead they made the 
reasonable-looking simplifications of (a) calculating phonons and
EP matrix elements for the undoped insulator, (b) moving the Fermi level in a rigid-band
fashion, and 
(c) using those quantities to evaluate $q$-dependent coupling ($\lambda_q$;
$q$ includes the branch index) and
finally $\lambda$, predicting T$_c$ up to 50K. 
In this paper we provide the resolution to this discrepancy, which involves 
the crucial effect of large doping-induced
spectral weight redistribution due to non-rigid-band shifts of spectral density
upon doping. The interlayer charge transfer underlying the shift in spectral density has
the same origin as the charge transfer obtained from alkali atom adlayers on
oxygenated\cite{CsODi} and native\cite{nativeDi} diamond surfaces to produce
negative electronic affinity structures. 
This ``mechanism'' of electronic structure modification will be useful in
designing materials for functionalities other than superconductivity. The spectral
shifts are distinct from those discussed in the doping of a Mott insulator
as we discuss below.

%\section{Methods and Structure}

First principles electronic structure calculations were performed within density functional 
theory (DFT) using the FPLO code\cite{FPLO} to obtain the electronic structure for both
undoped and doped materials, the latter one being carried out in the virtual crystal 
approximation (VCA), where the (say) Al$_{1-x}$Mg$_x$ sublattice (Ca substitution
is also an option) that gives up its valence
electrons is replaced by an atom with an averaged nuclear charge. VCA allows charge
transfer to be obtained self-consistently, neglecting only disorder in the Al-Mg layer. 
The result is the transfer of $x$ electrons per f.u. from Cu, with half going to
each of the neighboring Al-Mg layers,
corresponding to metallic Cu $d^{10-x}$.
Phonon spectra and electron-phonon coupling calculations for the doped system 
were performed using {\sc Abinit}\cite{abinit} version 6.6.3 with 
norm-conserving Trouiller-Martins pseudopotentials.
In both codes the Perdew-Wang 92 GGA (generalized gradient approximation) 
functional\cite{PerdewWang92} was used.
The phonon and EPC calculations were done on the rhombohedral unit cell
using a 24$^3$ k-point mesh and an 8$^3$ q-point mesh, interpolated to more q-points.

The measured structural parameters\cite{koehler} for CuAlO$_2$ used were for 
rhombohedral R$\bar{3}$m (\#166) structure
with $a = 5.927$ \AA, $\alpha = 27.932^\circ$. 
This structure is equivalent to $a = 2.861$ \AA, $c = 17.077$ \AA~with hexagonal axes. Cu 
resides on the $1a$ site at the origin, Al is at the $1b$ site, at 
($\frac{1}{2}$, $\frac{1}{2}$, $\frac{1}{2}$) and the O atom is in the 
$2c$ position ($u$, $u$, $u$), $u = 0.1101$.

%`\section{Electronic Character}

\begin{figure}[th]
\includegraphics[width=0.95\columnwidth,clip]{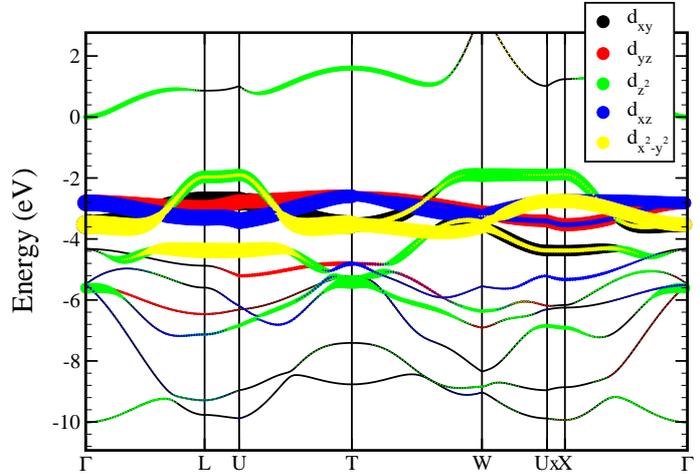}
\caption{(color online) Fatbands plot for CuAlO$_2$, with zero of energy at the top of the gap. 
The size of the symbol represents the 
amount of $3d$ character, and the color the character as given in the legend.}
\label{fig:band_weights}
\end{figure}

\begin{figure}[th]
\includegraphics[width=0.95\columnwidth,clip]{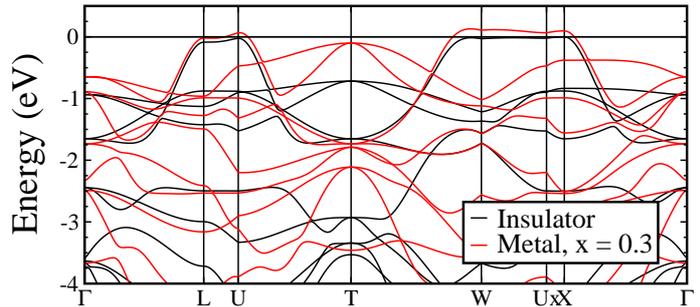}
\caption{(color online)Comparison of band structures for the metallic and 
insulating states of CuAl$_{1-x}$Mg${_x}$O$_2$ with $x = 0.3$. This moderate 
level of doping results in very strong changes in the relative band positions.}
\label{fig:bands_compare}
\end{figure}

\begin{figure}[th]
\includegraphics[width=0.95\columnwidth,clip]{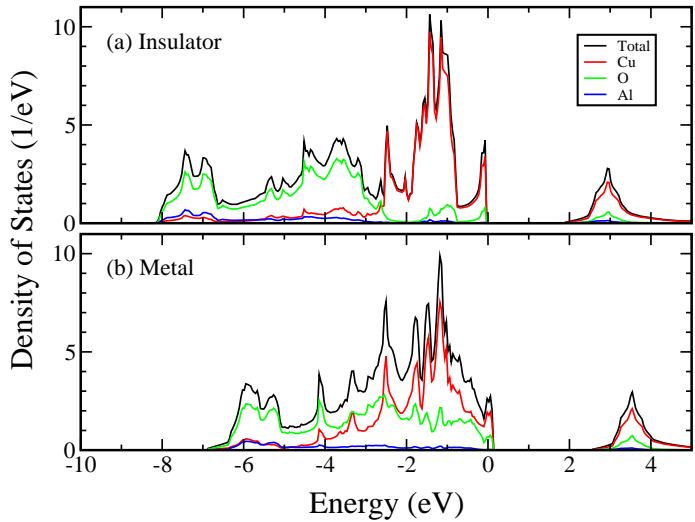}
\caption{(color online)Comparison of the density of states for the (a) insulating 
CuAlO$_2$ and (b) metallic CuAl$_{1-x}$Mg${_x}$O$_2$ with $x = 0.3$. For the insulator, 
the Cu $d$ bands are rather separate from the O $p$ bands, but upon doping
strong O $p$ permeates the Cu $d$ bands, to near the Fermi level.}
\label{fig:dos_compare}
\end{figure}

\begin{figure}[th]
\includegraphics[width=0.95\columnwidth,clip]{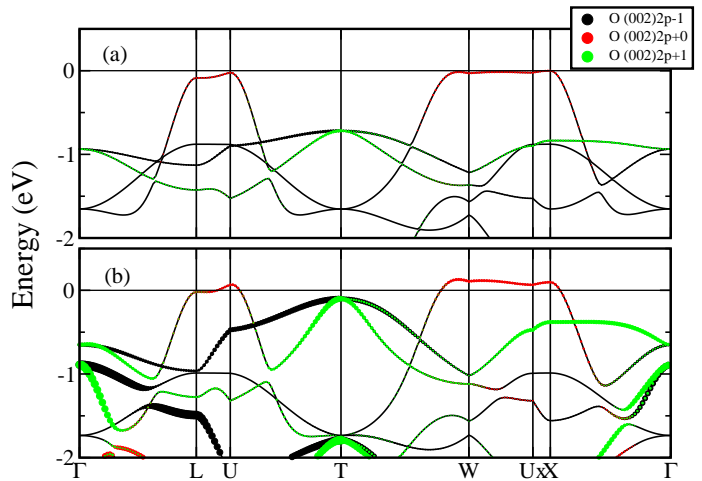}
\caption{(color online)Fatband plots for the (a) insulator and (b) $x$=0.3 metal states, emphasizing
O $2p$ character. In addition to the strong shift upward, the O $2p$ character has 
increased many-fold for the bands near E$_F$ in the metal. }
\label{fig:O_hybrid_compare}
\end{figure}

The band structure of insulating CuAlO$_2$ shown in Fig. \ref{fig:band_weights}, 
which agrees with previous work,\cite{NK-Y,AJF,Yanagi} 
illustrates that Cu $3d$ bands form a narrow, 2.5 eV wide complex at
the top of the valence bands. Oxygen $2p$ bands occupy the region -8 eV to -3 eV below the gap.
This compound is a closed shell
Cu$^+$Al$^{3+}$(O$^{2-}$)$_2$ ionic insulator with minor metal-O covalence, although
enough to stabilize this relatively unusual, strongly layered structure.

The upper valence bands providing the  hole states consist
of $d_{z^2}$ character with some in-plane 
$d_{xy}$, $d_{x^2-y^2}$ mixing.
The top of this band occurs at the edge of Brillouin zone (BZ) as in, for example, graphene,
but it is anomalously flat along the edge of the zone, viz. Ux-W (M-K, in hexagonal notation),
which comprises the entire edge of the BZ. 
Since it is also
almost dispersionless in the $\hat z$ direction, the resulting density of states 
just below the gap reflects a {\it one-dimensional phase space},
as shown in Fig. \ref{fig:dos_compare}a.  The $d_{xy}, d_{x^2-y^2}$ 
bands are nearly flat in the -2 to -1 eV
region, and  
the $d_{xz}, d_{yz}$ bands are even flatter, at -1 to -0.5 eV.
These four flat bands reflect very minor $d$-$d$ hopping in the plane.

When hole-doped, a dramatic shift of spectral weight occurs in the occupied bands, 
as is evident in both Figs. \ref{fig:bands_compare} for the bands and 
\ref{fig:dos_compare} for the spectral density.  With
the top (Cu $d_{z^2}$) conduction band as reference, the $3d$-$2p$ band complex at
all lower energies
readjusts rapidly with doping to lower binding energies. 
The $d_{xz}$, $d_{yz}$ bands (Fig. \ref{fig:band_weights}) 
acquire considerable $2p$ character and  move up to nearly touch E$_F$ 
at the point T=$(0,0,\pi/c)$; further doping will
introduce holes into this band.  The O $2p$ bands, which 
lay below the $3d$ bands in the insulator, have shifted upward dramatically by 2 eV
(a remarkably large 70 meV/\% doping), contributing extra screening
at and near E$_F$ in the metallic phase. The gap increases by $\sim$0.5 eV.
These spectral shifts can be
accounted for by a charge-dipole layer potential shift due to the Cu$\rightarrow$Al-Mg
layer charge transfer.
The increased $3d-2p$ hybridization is made more
apparent in Fig. \ref{fig:O_hybrid_compare}, which reveals that the $d_{xz}$, $d_{yz}$ 
bands at T (and elsewhere) have increased
contribution from the O $p$ states. Also apparent in this plot 
are seemingly extra bands appearing 
at about -1 eV 
near $\Gamma$; these are bands from below which have been shifted strongly upward by 
$\sim$2 eV by the dipole potential shift resulting from charge transfer. 

\begin{table}[bh]
\begin{centering}
\begin{tabular}{ccccc|ccc}

\hline \hline 
&	&	\multicolumn{3}{c|}{Insulator } & \multicolumn{3}{c}{Metal} \\
&	&	$z^2$& $xy$ & $x^2-y^2$ & $z^2$ & $xy$ & $x^2-y^2$ \\
\hline \hline 

\multirow{4}{*}{$z^2$}  
&	$t_1$		& \bf 393 &	198  &	228  &	\bf 342 &	191 & 220\\
&	$t_2$		&      60 &	8    &	13   &	     60 &	14  &  17\\
&	$t_3$		& \bf  35 &	22   &	25   &	\bf  59 &	17  &  20\\
&	$t_\perp$	& \bf  24 &	15   &	-16  &	\bf  63 &	17  &  17\\
\hline 
\multirow{4}{*}{$xy$} 
&	$t_1$		&	& 123 &	 117 &	& 107 & 107 \\
&	$t_2$		&	& 35  &	 14  &	&  36 &  16\\
&	$t_3$		&	& 11  &	 8   &	&  11 &  10\\
&	$t_\perp$	&	& 23  &	 14  &	&  31 &  20\\
\hline
\multirow{4}{*}{$x^2-y^2$} 
&	$t_1$		&	&  &	147 &	& & 140\\
&	$t_2$		&       &  &    28  &   & &  27\\
&	$t_3$		&	&  &	15  &	& &  17\\
&	$t_\perp$	&	&  &	18  &	& &  23\\
\hline

\end{tabular}
\caption[Tight binding parameters]
{Tight binding hopping parameters for insulating and metallic phases, 
from the three constructed Wannier functions.
The labels $t_1$, $t_2$, $t_3$, refer to the first, second, and third neighbor hoppings in the triangular Cu planes. 
$t_\perp$ refers to hopping between layers. The most significant 
changes when doped are highlighted in bold print.}
\label{tbl:TightBinding}
\end{centering}
\end{table}

\begin{figure}[th]
\begin{center}
%  \subfigure{
%  \includegraphics[width=0.45\columnwidth,clip]{WF3d0-ins.eps}
%  }
\subfigure{
\includegraphics[width=0.65\columnwidth,clip]{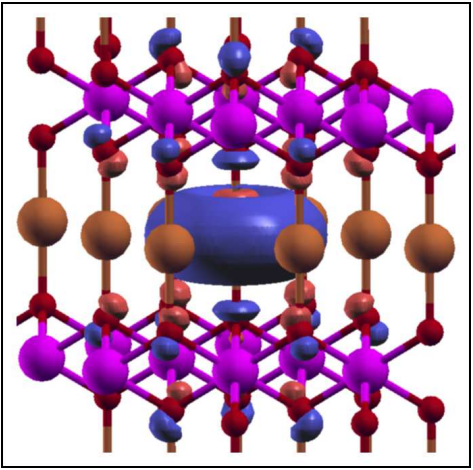}
}
\end{center}
\caption{(Color online.) Isosurface of the Wannier function for the Cu d$_{z^2}$ orbital in 
%  the (a) insulating and (b) metallic state. 
the $x$=0.3 doped metal. Antibonding contributions are seen from the nearest O 
atoms (small red spheres). The metallic state contains contributions from O ions in the second layer
above and below that are not present in the insulator.}
\label{fig:WF3d0}
\end{figure}

More light is shed on the electronic structure of CuAl$_{1-x}$Mg$_{x}$O$_2$ 
by using Wannier functions (WFs) to construct
a tight binding model of the states near the Fermi level. We use the 
WF generator in the FPLO code.\cite{FPLO} These WFs are symmetry 
respecting atom-based functions,\cite{weiku} constructed by 
projecting Kohn-Sham states onto, in this case, the Cu 3d$_{z^2}$, 3d$_{xy}$, 
3d$_{x^2-y^2}$ atomic orbitals, with resulting hopping amplitudes shown in 
Table \ref{tbl:TightBinding}. Hoppings involving the $xy$ and $x^2-y^2$ orbitals are not 
significantly different between the insulator and metal. However, hopping amplitudes 
for the $d_{z^2}$ WF change significantly, the most important being the factor of 2.5 
increase in the {\it hopping between layers}, $t_\perp$. Consistent with the picture from the DOS, 
the hoppings for the metallic state are more long-range: nearest neighbor hopping 
drops by 13\%, while third neighbor hopping nearly doubles. All of these changes
are neglected in a rigid band treatment.

This band dispersion is anomalous for a quasi-2D structure such as this, where normally the $3d$ orbitals
with lobes extending in the $x-y$ plane would be expected to be the most dispersive.  Instead, it is the
$d_{z^2}$ band that disperses, with a bandwidth of 2.5 eV and the band bottom at $\Gamma$. Shown in 
Fig. \ref{fig:WF3d0} is the $d_{z^2}$-projected WF for the $x$=0.3 hole-doped metal. 
Consistent with their minor dispersion, the WFs for the other $3d$ 
orbitals (not shown) have little contribution 
beyond the atomic orbital, showing only
minor anti-bonding contributions from nearby O atoms.  

The $d_{z^2}$ WF shape is, in addition, quite extraordinary. 
Although displaying $d_{z^2}$ symmetry as it must, its shape differs strikingly from atomic form.  It is so much
fatter in the $x$-$y$ plane than the bare $d_{z^2}$ orbital 
that it is difficult to see the signature $m_{\ell} = 0$ ``$z^2$'' lobes pictured in textbooks.
This shape is due, we think, to 
``pressure'' from the neighboring antibonding O $p_z$ orbitals above and below. 
There is an (expected) admixture of
O $2p_z$ orbitals, as well as a small symmetry-allowed $p_z + (p_x,p_y)$ contribution from the neighboring 
oxygen ions that finally provides (with their overlap)
the in-plane dispersion of the $d_{z^2}$ band. 
The important qualitative difference compared to the insulator WF is the
contribution from O atoms in the {\it next nearest} planes
(across the Al layer) whose states have been shifted upward by the
doping-induced charge transfer. This mixing opens a channel
for hopping between layers in the Cu $d_{z^2}$ WFs by creating overlap in the 
two planes of O atoms between Cu layers, it
is the source of the increase in $t_\perp$ hopping seen in Table \ref{tbl:TightBinding}
that leads to the $k_z$ dispersion of the $2p$ band along L-U in Fig. \ref{fig:O_hybrid_compare}
(and more so along $\Gamma$-T, not shown),
and will promote good hole-conduction in hole-doped delafossites.

\begin{figure}[th]
\centering
\subfigure{
\includegraphics[width=0.45\columnwidth,clip]{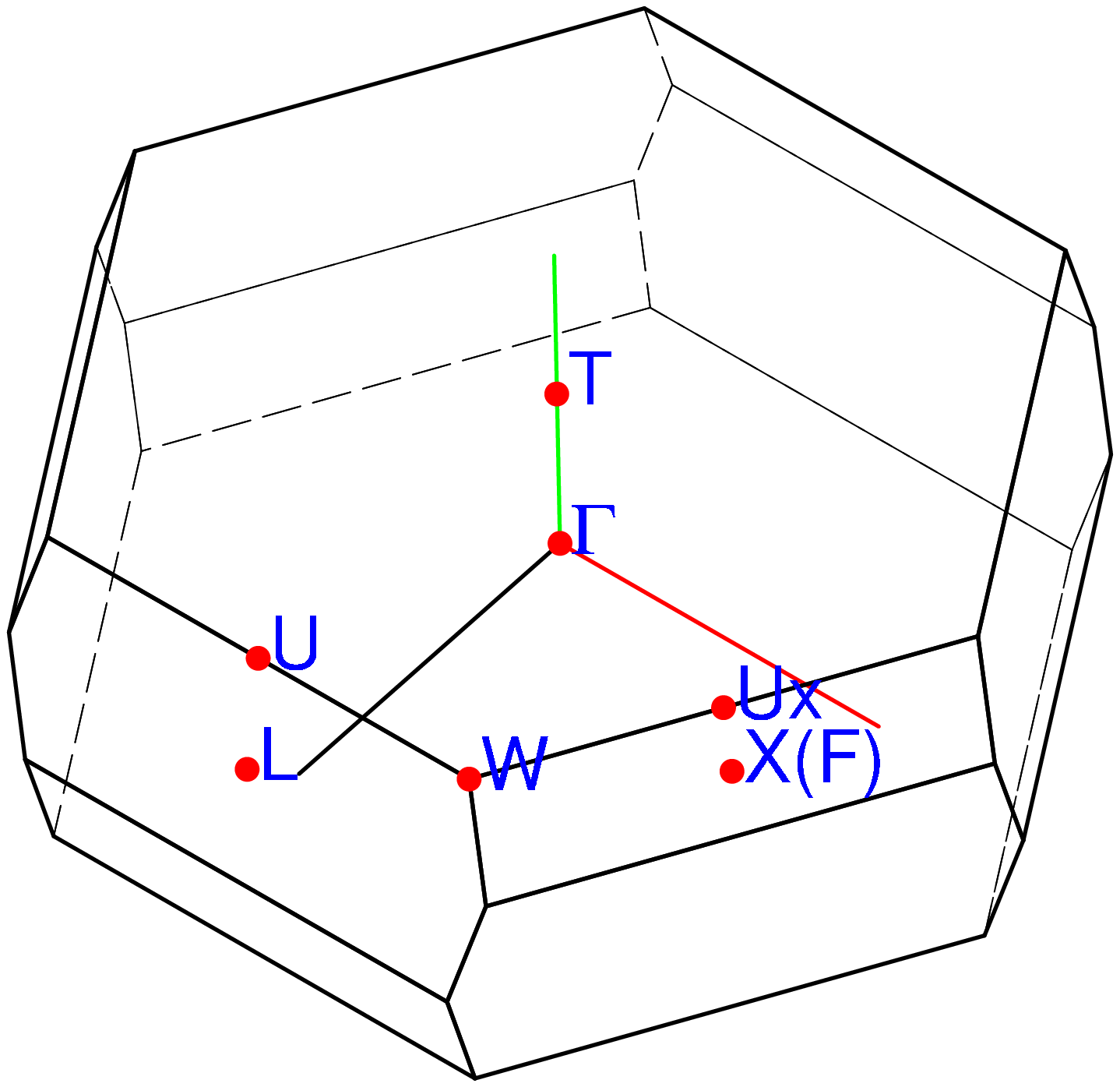}
\label{fig:kpoints}}
\subfigure{
\includegraphics[width=0.45\columnwidth,clip]{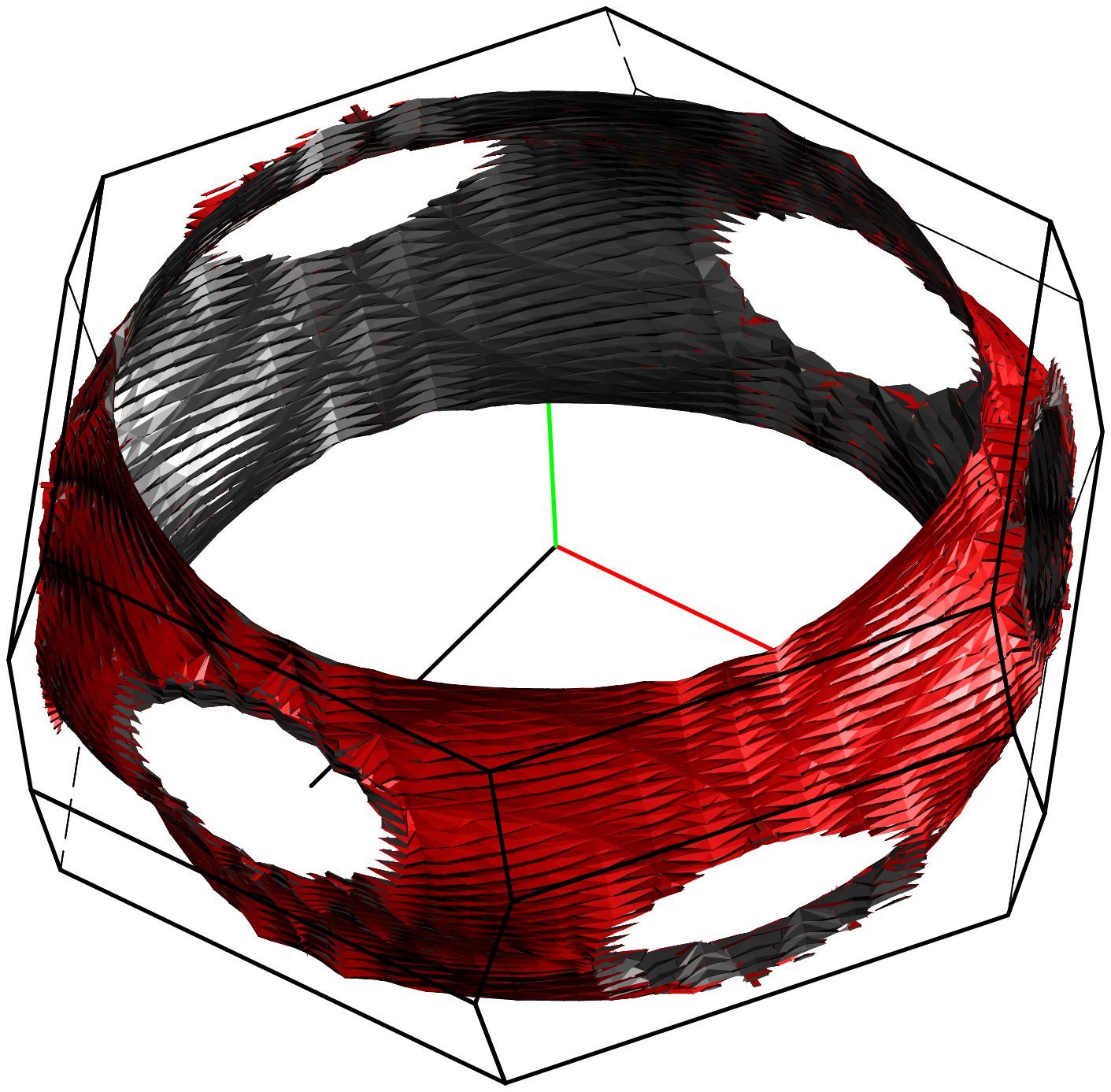}
\label{fig:Fermi}}
\caption{(color online)Left: rhombohedral zone with special k-points labeled.
Right: the sole large multiply-connected Fermi surface for moderately hole doped
CuAl$_{1-x}$Mg$_x$O$_2$, $x = 0.3$.}
\end{figure}

Fermi surfaces (FS) are critical to a material once it is doped into a metallic phase. For small
hole doping, the FS 
lies close to the zone boundary everywhere.
The FS of CuAlO$_2$ for $x=0.3$ hole doping in VCA,
displayed in Fig. \ref{fig:Fermi}, is not so different from that shown by NK for rigid band doping,
but the self-consistent treatment will differ substantially, 
for larger doping levels,
with new sheets appearing due to the spectral weight transfer.
The FS resembles a somewhat bloated cylinder truncated by the faces of the rhombohedral BZ.
The relevant nesting, not necessarily strong,
is of two types.  A large $2k_F$ spanning wavevector almost equal to the BZ dimension in the 
$k_x-k_y$ plane will, when reduced to the first BZ, lead to small $q$ scattering on the FS, broadened
somewhat by the $k_z$ dispersion.
Second, there are ``skipping'' $\vec q$ values along ($\epsilon,\epsilon$,$q_z$) for small $\epsilon$.
It is for these values of $\vec q$ that NK reported extremely strong coupling.
We have focused our study of EPC on the regime $x \sim$ 0.3 of doping where NK predicted the very large
electron-phonon coupling and high T$_c$.   

\begin{figure}[th]
\includegraphics[width=0.95\columnwidth,clip]{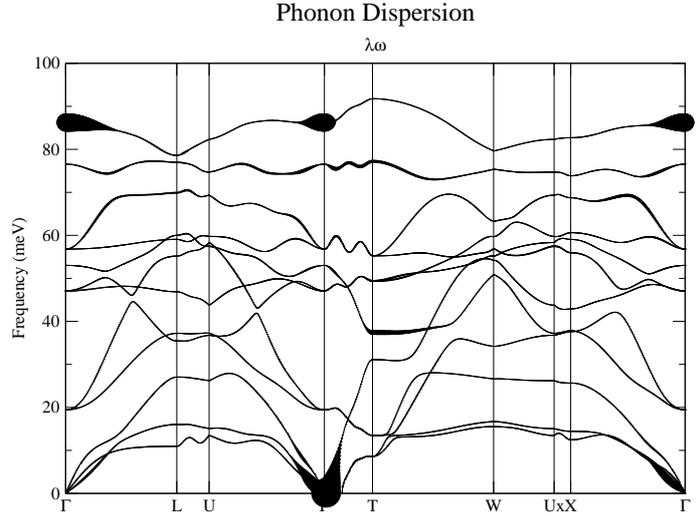}
\caption{Phonon dispersion curves for $x$=0.3 hole-doped CuAlO$_2$, calculated
with the {\sc Abinit} code on a
$8^3$ $q$-point grid with $24^3$ k-points. Circles indicate the magnitude of 
$\lambda_q \omega_q$ for that mode. Some aliasing effects (unphysical wiggles)
along L-U and $\Gamma$-T are due to the discrete nature and orientation
of the $q$-point mesh. 
}
\label{fig:lambda-omega}
\end{figure}

\begin{figure}[th]
\includegraphics[width=0.95\columnwidth,clip]{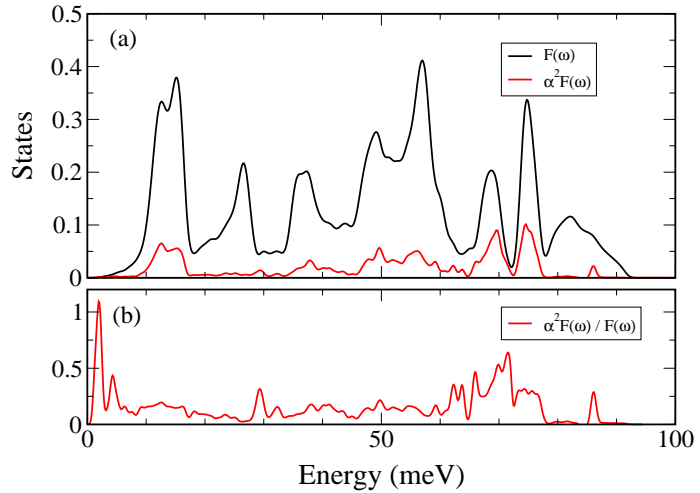}
\caption{(a) The phonon density of states and $\alpha^2F(\omega)$ at $x = 0.3$. 
(b) The quotient $\alpha^2(\omega)=\alpha^2F(\omega)/F(\omega)$ reflecting the 
spectral distribution of the coupling strength.
The peaks below 5 meV are numerically uncertain and are useless for EPC due to the vanishingly
small density of states.}
\label{fig:a2f}
\end{figure}

To assess the effects of the spectral shifts, we have computed the phonons and 
electron-phonon using linear response theory.
The phonon dispersion curves calculated from DFT linear response theory at $x$ = 0.3 are presented in 
Fig. \ref{fig:lambda-omega}, with fatbands weighting by $\omega_q \lambda_q$ (which is more
representative of contribution to T$_c$ than by weighting by the 
``mode-$\lambda$'' $\lambda_q$ alone\cite{PBA}). 
Branches are spread fairly uniformly over the 0-90 meV region. As found by NK, coupling strength
is confined to the Cu-O stretch mode at 87 meV very near $\Gamma$, and to very low
frequency acoustic modes also near $\Gamma$ where the density of states is very small.
Unlike in MgB$_2$, this coupling does {\it not} extend far along $k_z$; the lack of strong
electronic two-dimensionality degrades EPC coupling strength greatly and no modes
show significant renormalization.
We obtain $\lambda$ $\approx$ 0.2, $\omega_{log}$=275K = 24 meV.  
Using the weak coupling expression with $\mu^* \sim 0.1$ we obtain
\begin{eqnarray}
T_c \approx \frac{\omega_{log}}{1.2} e^{-\frac{1}{\lambda-\mu^*}}
   \sim 230 ~ e^{-10}K,
\end{eqnarray}
so no observable superconductivity is expected.
%Since $\lambda$ is so 
%close to the presumed value of
%retarded Coulomb repulsion $\mu^*$=0.10-0.13, the resulting critical temperature
%T$_c$ evaluated from the Allen-Dynes equation is vanishingly small.

% T_c = \frac{\omega_{log}}{1.2} \exp \left( -\frac{1.04(1+\lambda)}{\lambda - \mu^*(1+0.62\lambda)} \right)

Similar to that obtained by NK, the largest electron-phonon coupling arises from
the O-Cu-O bond stretch mode. As anticipated from the FS shape, 
the most prominent contributions arise from small $q$ phonons. 
The EPC spectral function
$\alpha^2F(\omega)$ is compared in Fig. \ref {fig:a2f}(a) with the phonon DOS $F(\omega)$. 
As is apparent from their ratio shown in Fig. \ref{fig:a2f}b, the peak around 
15 meV is purely from the large density of states there, due to the flat phonon bands 
over much of the zone at that energy.
The coupling with much impact on T$_c$ ({\it i.e.} area under $\alpha^2 F$)
occurs in the 45-75 meV range, and is spread
around the zone; however, unlike MgB$_2$ no frequency range is dominant and the coupling is weak. 
The top O-Cu-O stretch move, with the largest $\lambda$ values and in the 80-90 meV range,
are so strongly confined to narrow $q$ ranges that they contribute little to the coupling.

While we conclude, morosely, that high T$_c$ EPC superconductivity will not occur in doped CuAlO$_2$, 
the behavior that has been uncovered provides important insight into materials properties design
beginning from 2D insulators.  In the 40K electron-phonon superconductor MgB$_2$ superconductor,
an interlayer charge transfer of much smaller magnitude and natural origin self-dopes 
the boron honeycomb sublattice
to become the premier electron-phonon superconductor of the day. 
Hole doping of this delafossite does not provide better superconductivity, but it does
provide insight into designing materials behavior as well as providing a new platform for complex electronic
behavior. For low concentrations small polaron transport has been observed.\cite{AJF}
The hole-doping spectral shifts are distinct from doping-induced spectral shifts in
Mott insulators, which typically occurs without charge transfer.
As for the envisioned behavior: at moderate doping this materials class provides a single band (Cu
$d_{z^2}$) triangular lattice
system, with Cu$^{2+}$ S=1/2 holes, which if coupling is antiferromagnetic leads to frustrated magnetism.
The unusual dispersion at low doping, with little dispersion along $k_z$ and also around the zone
boundary, leads to an effectively {\it one dimensional phase space} at the band edge, although this
property degrades rapidly with doping.
Another triangular single band transition metal compound\cite{stacy,linbo3} 
is LiNbO$_2$, which superconducts around
5K when heavily hole doped\cite{stacy} and whose mechanism of pairing remains undecided.

%\section{Acknowledgments}
This work was supported by DOE SciDAC grant DE-FC02-06ER25794 and a collaborative effort with
the Energy Frontier Research Center {\it Center for Emergent Superconductivity}
through SciDAC-e grant
DE-FC02-06ER25777. W.E.P. acknowledges the hospitality of the Graphene Research Center at the
National University of Singapore where this manuscript was completed.

\end{document}